\documentstyle[psfig,aps,prl,amsfonts,amssymb]{revtex}

\begin{document}
\title{{\it a priori} Probability  that Two Qubits are Unentangled}
\author{Paul B. Slater}
\address{ISBER, University of
California, Santa Barbara, CA 93106-2150\\
e-mail: slater@itp.ucsb.edu,
FAX: (805) 893-7995}

\date{\today}

\draft
\maketitle
\vskip -0.1cm

\begin{abstract}
In a previous study (Slater, P. B. (2000) {\it Eur. Phys. J. B.} 
 17, 471-480), several
remarkably simple {\it exact} results were found, in certain specialized
$m$-dimensional scenarios ($m \leq 4$), 
for the {\it a priori} probability that a pair
of qubits is unentangled/separable.
The 
measure used was the volume element of the Bures
metric (identically one-fourth the statistical distinguishability 
[SD] metric).
Here, making use of a newly-developed (Euler angle) parameterization
of the $4\times 4$ density matrices of Tilma, Byrd and 
Sudarshan, we extend the analysis to the complete 
15-dimensional convex set ($C$) of 
{\it arbitrarily} paired qubits --- the {\it total} 
SD volume of which is known to
be ${ \pi^8 \over 1680} = {\pi^8 \over 2^{4} \cdot 3 \cdot 5 \cdot 7} 
\approx 5.64794$. Using advanced 
quasi-Monte Carlo procedures (scrambled Halton sequences) for numerical
integration in this high-dimensional space, we 
approximately (5.64851) reproduce that value, 
while obtaining an estimate of .416302 for the
SD volume of separable states.  We {\it conjecture}
 that this is but an 
approximation to ${\pi^6  \over 2310} 
={ \pi^6 \over 2 \cdot 3 \cdot 5 \cdot 7 \cdot 11}
\approx .416186$. The ratio of the two volumes,
${8 \over 11 \pi^2} \approx .0736881$, would then 
constitute the exact Bures/SD probability
of separability. The SD area of the 
{\it 14}-dimensional {\it boundary} of $C$ is 
${142 \pi^7 \over 12285} = {2 \cdot 71 \pi^{7} \over 3^{3} \cdot 5 \cdot 7 
\cdot 13} \approx 34.911$, 
while we obtain a numerical estimate of 1.75414 for the
 SD area of the boundary
of separable states.
\end{abstract}

\vspace{.1cm}

\pacs{PACS Numbers 03.67.-a, 03.65.Ud, 02.60.Jh, 02.40.Ky}
\hspace{1cm} {\bf{Key words:}} Qubits, $SU(4)$, density matrix, Bures
metric, statistical distinguishability metric, entanglement, Euler 
angle parameterization, negativity, concurrence, quasi-Monte Carlo, 
scrambled Halton sequences, numerical integration, isoperimetric inequalities

\vspace{.15cm}

\.Zyczkowski, Horodecki, Sanpera, and Lewenstein  
\cite{zycz}, giving various
``philosophical'', ``practical'' and ``physical'' motivations, were the first apparently 
to pose the question of ``how many entangled \ldots states there are in the
set of all quantum states''. 
In a sequel to
\cite{zycz}, \.Zyczkowski examined to what extent the choice of a measure
in the space of density matrices describing $m$-dimensional quantum sytems
affects conclusions regarding the relative frequency of entangled 
and unentangled (separable/classically correlated) states \cite{zycz2}.
Also, \.Zyczkowski and Sommers analyzed several product measures in the space of
mixed quantum states, in particular measures induced by the operation of
partial tracing  \cite{zycz3}.

In this general context, it was argued by Slater 
\cite{slater1,slater2} --- in 
analogy to the use classically in Bayesian theory 
of the volume element of the
Fisher information metric as Jeffreys' prior \cite{kass} --- that a 
natural measure on the quantum states would be the volume element of the 
Bures metric \cite{b1,b2,b3,slater4,jochen,kwek2}. 
``\ldots 
the Bures metric is locally equivalent to a Riemannian metric defined 
by the quantum analogue of the Fisher information 
matrix'' \cite[p.60]{holevo}.
``This metric provides a unitarily invariant measure for distinguishing
between two quantum states, and has been strongly motivated as physically
relevant both on measurement and statistical grounds'' \cite{mjw}.
Hall found  compelling evidence, at least in the case of 
the {\it two}-dimensional quantum systems, 
that the Bures metric induces the ``minimal-knowledge ensemble'' 
over the space of density matrices \cite{mjw}.
An additional distinguishing feature of the Bures metric is that the 
associated connection form (gauge field) satisfies the source-free
Yang-Mills equation \cite{dittYM,slaterYM}.
Chen, Fu, Ungar, and Zhao have interpreted the Bures fidelity between
possible states of a qubit in terms of the hyperbolic geometry applicable
to special relativity \cite{ungar}.
While the Bures metric fulfills the role of the {\it minimal} monotone
metric, there are a nondenumerable number of other monotone metrics as
well, all satisfying certain desirable statistical properties
\cite{petzsudar}.
It appears that all these other (non-minimal) monotone metrics would
lead to {\it lower} estimates of the proportion of states that are
separable/nonquantum in nature \cite{slater1}.
So, in this sense, the Bures metric provides upper bounds on
reasonable/acceptable estimates of separability.

In \cite{slater2}, specific use was made of the 
volume element of the Bures metric as a measure to address the question initially posed by \.Zyczkowski {\it et al} \cite{zycz}. This 
led to a number of quite surprisingly simple
probabilities of separability, 
such as ${1 \over 4}$ (Werner states), ${1 \over 2}$, $\sqrt{2}-1$ and 
${2 \over \pi} -{1 \over 2}$, when applied to
pairs of quantum bits (qubits) in certain restricted 
low-dimensional scenarios (cf. \cite{vidal}) 
for which exact integrations could be 
performed. (In several of these instances, the two {\it individual} qubits
were in the fully mixed or classical state and constraints were placed on 
possible correlations between the two qubits.)  In the present
 study, we seek to
remove any such limitations
and determine the Bures 
probability of separability of two qubits 
in the  full 15-dimensional framework.
However, due to the increased dimensionality/complexity, it appears
necessary (at least with the current state of development of
appropriate mathematical software) 
to have recourse to numerical methods for the requisite 
integrations. For this purpose, we rely upon recent developments in 
quasi-Monte Carlo procedures \cite{qmc} --- namely, the use of scrambled
Halton sequences \cite{faure,okten}. Upon the basis of our numerical results,
we formulate a 
conjecture (still awaiting {\it formal} proof or disproof), having 
interesting number-theoretic properties, 
that the Bures probability of separability of two 
arbitrarily paired qubits is
${8 \over 11 \pi^2} \approx .0736881$.

The joint state of two qubits is describable by a $4 \times 4$ density
matrix ($\rho$)--- that 
is, a Hermitian matrix, having trace 1 and nonnegative
eigenvalues. Such a state 
is separable (that is nonentangled or, equivalently, classically correlated)
if it can be expressed as the convex sum of 
tensor products
of pairs of 
$2 \times 2$ density matrices (which, in turn, represent the states of 
individual qubits) \cite{werner}. 
Ensembles of separable states, as well as
of {\it bound entangled} states can {\it not}
 be ``distilled'' to obtain pairs
in singlet 
(total spin 0) 
states for {\it quantum} information processing \cite{hor3,ved}.
Let us note that the Bures metric (and the related concept of
{\it fidelity}) has been an important instrument in the currently
widespread study of bipartite and multipartite quantum systems
\cite{Vedral,xiang,chen}.

As a practical matter, 
the question of what parameterization of the $4 \times 4$ density matrices
to employ is quite important for computational purposes.
In \cite{slater2}, we  used the 
``polarization matrix density technique'' 
parameterization (based on tensor products 
of Pauli matrices) \cite[eq. (1)]{Mkr} \cite{Fano}, focusing on 
certain $m$-dimensional subsets ($m \leq 4$) of the 
fifteen-dimensional 
space of $4 \times 4$ density matrices. 
We were then able to obtain, as already indicated,
several simple {\it exact} values for the Bures probabilities of 
separability of two qubits, the possible joint states of which were restricted
to these low-dimensional spaces.   

In our initial study on the question of relative separability/entanglement
\cite{slater1}, we had relied upon the {\it naive} 
parameterization --- $\rho_{ij} = a_{ij} + i b_{ij}$, 
where the $a$'s and $b$'s are real --- of the 
15-dimensional convex set of 
$4 \times 4$ density matrices, in order to obtain a 
{\it number} of
estimates of the full, general
Bures 
probability (as well as the corresponding probability of separability 
for the 35-dimensional convex set of 
$6 \times 6$ density matrices, representing the joint state
of a qubit and {\it qutrit}). Because no analytical expressions
are known for the 
highly complicated 
boundary of the domain using this 
parameterization (cf. \cite{bloore}) (as well as for the polarization
density matrix technique),
it was necessary in \cite{slater1} to reject many points 
of the imposed lattices used for numerical 
sampling
since  they
turned out to be incompatible with the {\it positivity}
 requirement for density
matrices. Additionally, 
diagonal entries ($a_{ii}$) --- because they all sum to 
unity --- had to
be sampled differently (that is, from a probability simplex rather than 
a regular lattice) 
than the off-diagonal entries ($a_{ij},b_{ij}, i \neq j$). 
This led us to
report {\it several} estimates, each  
depending upon the particular resolutions 
used for selecting 
candidate diagonal and off-diagonal entries 
of the $4 \times 4$ density matrices \cite[Tables 1-3]{slater1}.
Most of the resultant estimates of the Bures probability of separability
were in the neighborhood of .1.
(For an analogous study of the Gaussian two-party quantum states,
see \cite{slatergauss}.)

In contrast to this somewhat nonideal situation, 
the recently-reported Euler angle parameterization of Tilma, Byrd
and Sudarshan \cite{tilma} (based on the diagonalization
$\rho= U \Lambda U^{\dagger}$, where $U$ is unitary) appeared to yield
 a domain --- relatively easy to numerically
integrate over --- that is 
simply
a 15-dimensional hyperrectangle. 
However, there was an erroneous claim made in \cite{tilma} regarding this,
requiring
rectification before we can proceed correctly. 
It was  stated that for the choice of ranges of the three spherical angles 
\cite[eq. (36)]{tilma} (the other twelve variables being the Euler angles
parameterizing the unitary matrix $U$ drawn from the 
Lie algebra $SU(4)/Z(4)$),
\begin{equation} \label{rectdom}
{\pi \over 4} \leq \theta_{1} \leq {\pi \over 2}; \quad
\cos^{-1}{1 \over \sqrt{3}} \leq \theta_{2} \leq
{\pi \over 2}; \quad 
{\pi \over 3} \leq \theta_{3} \leq {\pi \over 2},
\end{equation}
the vector of (nonnegative) eigenvalues,
\begin{equation} \label{tbs1}
(\lambda_{1},\lambda_{2},\lambda_{3},1-\lambda_{1}-\lambda_{2}-\lambda_{3}) 
=(\sin^{2}{\theta_{1}} \sin^{2}{\theta_{2}} \sin^{2}{\theta_{3}},
\cos^{2}{\theta_{1}} \sin^{2}{\theta_{2}} \sin^{2}{\theta_{3}},
\cos^{2}{\theta_{2}} \sin^{2}{\theta_{3}},\cos^{2}{\theta_{3}}),
\end{equation}
would be strictly ordered, that is, 
$\lambda_{1} \geq \lambda_{2} \geq \lambda_{3} \geq 1 
-\lambda_{1} -\lambda_{2} -\lambda_{3}$.
However, simple testing of ours revealed that while $\lambda_{1}$ is, in 
fact, always
at least as great as the other three eigenvalues, 
these last three do not necessarily conform
to any particular order within the ranges designated.
(We were not yet 
aware of this difficulty in our earlier study \cite{slatergoof},
and were led to erroneously assert there that the desired
Bures/SD probability of separability of arbitrarily paired qubits was
$\sqrt{2} /24 \approx .0589256$.)

In \cite{slaterhall} we had specifically addressed
the question of generating an {\it ordered}
 vector (lying in the unit simplex).
In terms of the parameterization (\ref{tbs1}), these conditions can be 
expressed as
\begin{equation} \label{constr}
 {\pi \over 4}  \leq \theta_{1} \leq {\pi \over 2}; \quad
f(\theta_{1}) \leq \theta_{2} \leq {\pi \over 2};  \quad 
f(\theta_{2})  \leq \theta_{3} \leq {\pi \over 2};   \quad f(x) = 
{\cot^{-1} \Big( {\cos{{x}}}} \Big).
\end{equation}
The Lebesgue measure of the Euclidean domain defined by the three-dimensional
volume determined
 by the ranges (\ref{constr}) is .0564221, while that defined 
by the ranges (\ref{rectdom}) is 4.48593 times larger, that is
.253106.

We could proceed further  with these ranges (\ref{constr}), but in order 
to avoid loss of the convenient hyperrectangular structure 
posited in \cite{tilma}, we have chosen
to simply employ
\begin{equation} \label{constr1}
0 \leq \theta_{1}, \theta_{2}, \theta_{3} \leq {\pi \over 2}.
\end{equation}
(The Lebesgue measure of which is, of course, 
 $({\pi \over 2})^3 \approx 3.87578$.)
These ranges generate all possible four-vectors (\ref{tbs1}) in
a unique manner.
 (Integrals using the
ranges (\ref{constr1}) would then, in the context here, simply be 
$4!=24$ times those obtained using the set of angular ranges (\ref{constr}).)

Given the (restored/modified) hyperrectangular structure, it is simple to 
rescale
each of the fifteen coordinates, so as to obtain a hyper{\it cubic} domain, 
with all edges equal to unity in length.
Most available quasi-Monte Carlo computer routines are written with
such a regular framework in mind \cite{qmc}.

Having so converted to the unit hypercubic structure, we placed 65 million
 (``low discrepancy'') points
over the hypercube, devised so as to be near to {\it uniformly}
distributed. The specific method employed was that of scrambled
Halton sequences \cite{faure,okten}. One of the classical low-discrepancy 
sequences is the van der Corput sequence in base $b$, where $b$ is any 
integer greater than one. The uniformity of the van der Corput numbers can be
further improved by permuting/scrambling
 the coefficients in the digit expansion
of $m$ in base $b$. The scrambled Halton sequence in $m$-dimensions --- which we employ --- is 
constructed using
the so-scrambled van der Corput numbers for $b$'s ranging over 
the first $m$ prime numbers \cite[p. 53]{okten}.

Let us now discuss an issue largely of 
terminology, but important to keep in mind
in implementing various formulas. 
Braunstein and Caves \cite{b3} showed that the Bures metric
(as stipulated in \cite{b1,b2}) was equal to identically
{\it one-quarter} of their statistical distinguishability (SD) metric 
(cf. \cite[eq. (2.29)]{mjw}).
However, Hall \cite{mjw}, 
citing \cite{b3}, spoke in terms of the Bures
metric, but actually employed the formulas for the SD 
metric \cite[eq. (24)]{mjw}.
This, of course, is a matter of no consequence if one computes weighted
averages or probabilities with respect to the volume element of one metric 
or another. It is pertinent, however, when absolute rather than relative 
volumes
are to be determined, with a factor of $4^{m (m-1) \over 2}$ difference
occurring for $m$-level systems.
Thus, the Bures volumes themselves will be 
 $4^{-6}$ times smaller than
the SD ones (of a somewhat more appealing form) given below for our case
of $m=4$.

We computed the corresponding statistical distinguishability (SD) volume 
element (``quantum Jeffrey's prior''
\cite{kwek2}) at each point of the scrambled Halton sequence in 15 dimensions. 
This volume element is 
the product of the {\it Haar} volume element over the
{\it twelve} Euler angles parameterizing $SU(4)/Z(4)$
\cite[eqs. (24), (25)]{tilma} and 
the {\it conditional} SD volume element
 over the {\it three}-dimensional simplex of eigenvalues
\cite[eqs. (16), (17)]{slaterhall}. 

The {\it conditional} SD volume element ($\mbox{d}D_{n}$) 
over an $(m-1)$-dimensional 
simplex
of (nonnegative) eigenvalues, constrained to sum to 1, 
 can be expressed as (cf. \cite[eq. (24)]{mjw}),
\begin{equation} \label{volelement}
\mbox{d}D_{m} = 
{\mbox{d} \lambda_{1} \ldots \mbox{d} \lambda_{m-1}  \over 
\sqrt{ \Pi_{i=1}^{m} \lambda_{i}}}
 {\Pi_{1 \leq i <j}^{m} {4 (\lambda_{i} -\lambda_{j})^2 \over 
(\lambda_{i}+ \lambda_{j})}}.
\end{equation}
(If the factor of 4 is omitted, this becomes the conditional Bures volume
element.)
Integrating (\ref{volelement}) over the simplices for various $m$,
we obtain, as far as we have been able to compute exactly, 
\begin{equation}
D_{2} = 2 \pi \approx 6.28319; \quad  D_{3} = {64 \pi \over 35} = 
{2^6 \pi \over 5 \cdot 7}
\approx 5.74463; \quad D_{4} = {2  \pi^2 \over 35} ={2 
\pi^{2} \over 5 \cdot 7} \approx .563977;
\end{equation}
 \begin{displaymath} 
D_{5} = {8388608 \pi^2 \over 156165009} ={2^{23} \pi^2 \over 
3 \cdot 7^{2} \cdot 11 \cdot 13 \cdot 17 \cdot 19 \cdot 23} 
\approx .530159.
\end{displaymath}
(Also, $D_{6} \approx {4^{15}  \pi^3 \over 1.53636 \cdot 10^{16}} 
\approx 2.16436 10^{-6}$.)

The ``truncated'' Haar volume of the Lie algebra $SU(4)/Z(4)$, using the 
Euler angle parameterization \cite{tilma}, is ${\pi^6 \over  96}$
\cite{tilma}, 
so the total 
(separable plus nonseparable) SD volume ($V^{s+n}$) 
of the 
15-dimensional convex set of four-level quantum systems
 is the {\it product} of this term 
and the term $D_{4}$ \cite[eq. (24)]{mjw},
\begin{equation} \label{sn}
V^{s+n} =  {\pi^6 \over 96} \cdot {2 \pi^2 \over 35 } = 
{ \pi^{8} \over 1680 } = 
{ \pi^{8} \over 2^{4} \cdot 3 \cdot 5 \cdot 7} \approx 5.64794.
\end{equation}
``Truncation'' occurs because three of the fifteen Euler angles, 
corresponding to diagonal Lie generators, become irrelevant (that is,
``drop out'') in the formation of $\rho$. Without truncation, 
the appropriate Haar volume would equal ${\pi^9 \over 288 \sqrt{2}}$
\cite[eq. (B24)]{tilma}.

We also 
determined whether or not the density matrix corresponding to each of the
65 million
points of the scrambled Halton sequences was separable.
 (``Essentially, the mathematical context is one of two nested 
compact convex sets and the determination of whether a point 
in the larger set is in the smaller set'' \cite{pittenger}.)
For this purpose, we employed the  {\it partial transposition} criterion of
Peres \cite{peres} and the  Horodecki trio \cite{Horodecki}.
(The partial transpose of a $4 \times 4$ density matrix can be obtained
by transposing in place each of its four $2 \times 2$ blocks.)
In fact, since no more than one eigenvalue of the partial transpose
of a 
 $4 \times 4$ density matrix
can be
 negative  \cite[Thm. 5]{frank0} \cite{tilma,sanpera}, one could
simply employ the sign of the determinant of the
partial transpose as the test for separability, rather than the positivity
of all four eigenvalues (cf. \cite{wang}). That is, 
a positive determinant of the partial
transpose informs us that the density matrix which has been partially
transposed is separable in nature, while a negative determinant tells us
it is nonseparable or entangled.

For the scrambled Halton sequence of 65 million points, distributed in a 
near-to-uniform manner
over the 15-dimensional unit hypercube, we obtained 5.64851 as an estimate of the (known)
total Bures volume (5.64794) 
and .416302 for an estimate of the (unknown) Bures volume of the
separable states. (For the initial 10 million points, the analogous
figures were 5.64615 and .415716, and for the initial 20 million,
5.64829 and .416775.)

Multiplying $V^{s+n}$  by 
the {\it probability} ${8 \over 11 \pi^2} \approx .0736881$, 
we obtain what --- on
the basis of our numerical evidence plus considerations of mathematical
simplicity/elegance, buttressed by our previous findings of
simple exact solutions in low-dimensional settings
\cite{slater2} ---  we {\it conjecture} 
to be the  SD volume of the {\it separable} 
$4 \times 4$ density matrices, that is,
\begin{equation} \label{conjsep}
V^{s} = {8 \over 11 \pi^2} 
V^{s+n} = 
{\pi^{6} \over 2310} = 
{ \pi^{6} \over  2 \cdot 3 \cdot 5 \cdot 7 \cdot 11} ={\pi^6 \over 11\#}
\approx .416186.
\end{equation}
The notation $p\#$ denotes the products of the primes less than or equal
to $p$ \cite{caldwell}, so $V^{s+n}$ can be expressed as
 ${\pi^8 \over 2^3 \cdot 7\#}$. Let us point out that the pair 
of integers (1680, 2310),
occurring in the denominators of $V^{s+n}$ and $V^{s}$, are the {\it last} two
members of a certain sequence (denoted A064377) of number-theoretic interest
\cite{sloane}, in that it
 has been conjectured by E. L\'abos \cite{sloane} that 1680 and 2310 
are the two largest integers for which the sum of the {\it fourth}
power of their divisors exceeds the {\it fifth} power of the number
(Euler's totient function) of positive integers relatively prime to them.

Since our numerical (quasi-Monte Carlo) estimate of $V^{s+n}$, that is 
$5.64851 > 5.64794$, errs on the
positive side, it is rather natural also to expect that our 
numerical estimate
of $V^{s}$ would err in the same direction. Subject to the validity of our
conjecture (\ref{conjsep}), this is, in fact, the case, since
$.416302 > .416186$.

In the computational process described above, we also determined the
average Bures/SD 
entanglement of each of the 65 million density matrices generated, using
two possible measures --- the {\it negativity} and the 
{\it concurrence}, the former
always being no greater than the latter \cite{frank}. For the mean 
negativity we obtained .177162 and for the mean concurrence,
.197284 (cf. \cite[Figs. 4(b), 5(a)]{zycz3}).
(Another interesting measure of entanglement to similarly study would be
 the Bures distance of a
quantum state to the separable states \cite{Vedral} 
(cf. \cite{Bertlmann,pittenger,belges}).)

Since the Bures/SD  probability of separability conjectured here 
(${8 \over 11 \pi^2} \approx .0736881$) of a pair of
qubits is somewhat less in value than those estimates
($\approx .1$) arrived at previously in
\cite{slater1}, using the naive parameterization 
($\rho_{ij} = a_{ij} + i b_{ij}$) of the $4 \times 4$ 
density matrices, 
we are led to assert that the ``world is even more quantum''
\cite{zycz} than previously indicated. In \cite{slater1} it appeared that
the Bures probability of separability provides a natural {\it upper}
 bound
on an entire class (corresponding to the monotone metrics \cite{petzsudar})
of possible prior probability measures. (The Bures metric serves as the
{\it minimal} monotone metric \cite{petzsudar}.)
On a qualitative level, then, knowledge that the joint state of two 
qubits is separable would be considerably 
more ``informative'' --- in allowing us to
estimate the underlying parameters of the state --- than knowledge 
that it is
entangled.

In a study \cite{slater4}
of the {\it eight}-dimensional convex set of the $3 \times 3$ 
density matrices, we developed explicit formulas
for the entries ($g_{ij}$) of the Bures metric tensor, based on the Euler
angle parameterization of $SU(3)$ \cite{mark1}. Several pairs 
($i,j$) of the eight
variables there were found to be mutually orthogonal (that is, $g_{ij}=0,  
i \neq j$).
We have examined the question of mutual orthogonality also in the present
$SU(4)$ case 
\cite{tilma} (but have not yet attempted to obtain 
simplified explicit formulas
for the $g_{ij}$'s, in general).
Our conclusions based on 
strong {\it numerical} evidence  --- obtained by 
implementing the ``explicit formulae for the Bures metric'' 
of Dittmann \cite{jochen} --- are that:
the twelve Euler angles 
$\alpha$'s (in the notation of \cite{tilma}) 
are each mutually orthogonal to the three 
(mutually orthogonal) $\theta$'s, parameterizing the eigenvalues
(\ref{tbs1});
$\alpha_{6}$ is orthogonal to $\alpha_{5}$ and also to all
$\alpha_{i}, i>6$; $\alpha_{9}$ 
is orthogonal to both $\alpha_{10}$ and
$\alpha_{12}$; and $\alpha_{10}, \alpha_{11}, \alpha_{12}$ are all
mutually orthogonal.
In the important
 inverse matrix ($||g_{ij}||^{-1}$) the twelve $\alpha$'s are again, 
obviously, 
all orthogonal with the three $\theta$'s, but no other such 
pairs were found (unlike the SU(3) case \cite{slater4,mark1}).

 Additionally, of course, we would like 
 to study in similar ways {\it higher}-dimensional 
bipartite and multipartite
quantum systems than that examined here.
Tilma and Sudarshan have given, along
with other higher-dimensional systems,
Euler angle-based parameterizations of the $8 \times 8$ 
density matrices of 
{\it three} qubits \cite{ts}, and 
indicated to the author the parameterization for the $6 \times 6$ density
matrices, corresponding to 
 a paired qubit and  {\it qutrit}.
(It would be desirable for any such analyses 
to know beforehand the precise conditional SD volumes, $D_{m}$,
 $m>5$, seeing that knowledge of $D_{4}$ was crucial in our
being able to formulate the conjecture as to the Bures volume of
separable $4 \times 4$ density matrices.)
We recall that for $m > 6$, the 
Peres-Horodecki 
partial transposition 
criterion provides a necessary, but not sufficient
condition for separability \cite{hor3,peres}.
As the Hilbert space dimensions of coupled $l$ and $m$-dimensional quantum
systems {\it increase}, we expect the 
corresponding Bures/SD probability of separability to
{\it decrease} \cite{zycz,slater1}.

Let us apply the formula (\ref{volelement}) for the conditional SD volume
$\mbox{d} D_{m}$, with $m=4$, but now setting, say, $\lambda_{1} 
\rightarrow 0$
(by taking $\theta_{1} \rightarrow 0$). 
Then, integrating the resulting expression, to high accuracy,
 over the 2-dimensional 
simplex, we obtain .871513859457. Multiplying this by a factor of four
(to account for the possibility that $\lambda_{2}, \lambda_{3}, \lambda_{4}$
or $1-\lambda_{1}-\lambda_{2}-\lambda_{3}$ 
is the zero eigenvalue) and then
by the truncated Haar volume, ${\pi^6 \over 96}$, we get 34.9110002722. 
This is the {\it 14}-dimensional SD surface {\it area}
($A^{s+n}$) 
of the boundary ($|\rho|=0$) 
of the 15-dimensional convex set of $4 \times 4$ density
matrices. It appears overwhelmingly 
 convincing that
\begin{equation}
A^{s+n} = {142 \pi^{7} \over 12285} = {2 \cdot 71 \pi^{7} 
\over 3^3 \cdot 5 \cdot 7 \cdot 13} \approx  34.91100027222665.
\end{equation}
(The denominator 12,285 is the number of permutations of 15 items
in which exactly 4 of them 
change places \cite[seq. A060008]{sloane}.) For the $3 \times 3$ density matrices, the SD area of the boundary
$|\rho| =0$ is $3 (512/63) (\pi^3 /2) = 256 \pi^{3} /21 \approx 377.981$, 
and for the $5 \times 5$ density matrices, $5 (.00736276442200) 
(\pi^{10} /18432) \approx .187041154554$. Note that $2439209213 \pi
/ 5716630 \approx .00736276442200$ with $5716630 = 2 \cdot 5 \cdot 41 
\cdot 73 \cdot 191$ and $2439209213 = 7 \cdot 348458459$.
For the $6 \times 6$ density matrices --- forming a 
35-dimensional space --- we have
$6 (3.85759  \cdot 10^{-6}) (\pi^{15} / 35389440) \approx .00001874312$.
Observe that $168 \pi^2 /v \approx .38548 \cdot 10^{-6}$, where
$v=430,137,400$ is the number of permutations of 35 items in which
exactly 6 of them change places. (For the simplest case of the $2 \times
2$ density matrices, we get $2 \pi^{2}$ for the SD volume and
$16 \pi$ for the SD area of the pure state boundary.)

We have also estimated $A^{s}$ itself --- based on
the first 11,800,000 points given by scrambled Halton sequences, now in
14-dimensional space. Of these 
11,800,000, we found 8,083,953  of them for which 
there existed at least one acceptable value (that is, lying between 0 and
1) 
of the 15-th coordinate 
(taken to be the rescaled form of $\theta_{3}$) for which the 
corresponding density matrix lay on the separable-nonseparable boundary (as 
indicated by a zero determinant of its partial transpose).
(DiVincenzo, Terhal, and Thapliyal considered situations is which a 
mixed state is ``marginally separable, in the first case because the partial
transpose of the state has zero eigenvalues, and in the second because the state is defined as the complement (in a larger Hilbert space) 
of a barely completable product
basis'' \cite{dave}.)
In total, we obtained 15,330,369  such (boundary) density matrices.
We then determined the associated SD area elements (identically 
$\pi^{-1}$  times in value the corresponding SD volume elements).
The  computations gave an estimate of $A^{s} \approx 1.75414$.
(Based on just the first 3,200,000  points of this sequence, 
the estimate of $A^{s}$ was 1.74893.)
If the SD area does, in fact, have a simple exact expression, it might
possibly be
\begin{equation}
A^{s} = {\pi^5 \over 175} = {\pi^5 \over 5^{2} \cdot 7} \approx 1.74868.
\end{equation}
(But also $\pi^6 /548 \approx  1.75436$.)

It would be of interest to study
the relations between $A^{s+n}$ and $V^{s+n}$, as well as between
$A^{s}$ and $V^{s}$, 
 in terms of isoperimetric
inequalities \cite{chavel,morgan,druet1,druet2} --- taking into 
account known curvature 
properties of the Bures metric \cite{ditty1,ditty2,ditty3} (cf. \cite{volper}). The scalar curvature of the 
Bures metric on a $4 \times 4$ density
matrix ($\rho$) has been expressed as \cite{ditty2},
\begin{equation} \label{SC}
S^{1} = 6 {63 e_{4} + 35 e_{3}^{2} -43 e_{2} e_{3} -7 e_{3} -3 e_{2}^{2} \over e_{4} +e_{3}^{2} -e_{2} e_{3}},
\end{equation}
where $e_{i}$ is the elementary invariant of degree $i$ of $\rho$ 
(that is, $\Pi_{i=1}^{m} 
(\lambda_{i} -t) =\sum_{i=0}^{m} e_{m-i} (-t)^{i}$), so that the
scalar curvature depends only on the eigenvalues ($\lambda$'s) of $\rho$. It is unbounded for $m>2$ and achieves its minimum, $(5 m^2 -4) (m^2-1)/2$, at 
the fully mixed or classical state, having the $m \times m$ 
density matrix $\rho ={1 \over m} \bf{1}$, 
so this minimum is  570 for $m=4$ \cite{ditty2}. (The sectional curvature
is also always greater than 1 \cite[eq. (6.2)]{ditlie}.) Also,
the $(m^2-1)$-dimensional space of $m \times m$ density matrices, representing the $m$-level quantum systems, is
{\it not} locally symmetric for $m>2$ \cite{ditty1}.
(We might also observe that a Euclidean 15-sphere 
(having a radius of 1.19682) with volume
equal to $V^{s+n}$ has a surface area of 70.7865 (cf. 34.911), while 
such a 15-sphere 
(having a radius of 1.01128) with
volume equal to $V^{s}$ has an area 6.20661 (cf. 1.74893).)

If, in addition, to the scalar curvature (\ref{SC}), 
the Ricci curvature were also
bounded below, in particular, by \newline $(m^2-1)-1=14$, 
then we could 
directly apply the
``Levy-Gromov'' isoperimetric inequality \cite[p. 520]{gromov}.
In this case, the ratio (.318581) of the area $A^{s} \approx 
1.75414$ of the 
boundary
of separable states to the volume $V^{s+n} =\pi^8/1680$
 would be {\it greater} than
the ratio ($w$) of $s(\alpha)$ to the volume $\tilde{V} =
256 \pi^7/2027025$ of
a unit ball in 15-dimensional space. Now $\alpha$ itself is the
ratio $V^{s} /V^{s+n}= 8 /11 \pi^2$ and $s(\alpha)$ is the area
of the boundary of a ball in 15-dimensional space having a volume
equal to $\alpha \tilde{V}$. This gives us $w=1.31521$, so the
Levy-Gromov 
inequality fails, since $.318581 \ngeq 1.31521$ and we are left to conclude that the Ricci curvature
for the qubit-qubit states endowed with the Bures metric must somewhere
assume a value less than 14. (However, for the qubit-qutrit case
\cite{slaterpreprint2}, no contradiction with the 
corresponding inequality appears to
hold.) To continue along these lines, there is an extension 
\cite[Thm. 6.6]{chavel2} of
the Levy-Gromov result from the case where the Ricci curvature is bounded
below not simply by $m^2-1$ but by $(m^2-1) \kappa$ (where 
$\kappa$ is interpreted as 
the constant sectional curvature of a 15-dimensional sphere 
of radius $1/\sqrt{\kappa}$).
Then, the corresponding isoperimetric inequality would not be inconsistent
with our particular values of $V^{s}, V^{s+n}$ and $A^{s}$ 
if the Ricci curvature were bounded below by 
 $\approx .780703$ (but nothing higher).

\acknowledgments

I would like to express appreciation to the 
Kavli Institute for Theoretical
Physics for computational  support in this research, 
to Mark Byrd and Karol \.Zyczkowski
for encouragement at various stages of this research program, and to 
Giray \"Okten for making available his MATHEMATICA program 
\cite{okten} for computing
scrambled Halton sequences.
(\"Okten 
 has also recently prepared
 software for scrambled {\it Faure} sequences, which may yield more
accurate estimates, as well as statistical error analyses.)
The final revision of the 
manuscript was undertaken
 at King Fahd University of Petroleum
and Minerals with the kind technical assistance of Mian Zainulabadin
 Khurrum of the
Information Technology Center there.

\end{document}